\documentclass{elsart}
\usepackage[dvips]{graphicx}

\usepackage{amssymb}

\begin{document}

\begin{frontmatter}



\title{Water adsorption and dissociation on BeO $(001)$ and $(100)$ surfaces}


\author[MHC]{Maria A. Gomez}
\address[MHC]{Department of Chemistry, Mount Holyoke College, 50 College Street, South Hadley, MA 01075}

\author[LANL]{Lawrence R. Pratt and Joel D. Kress}
\address[LANL]{Theoretical Division, Los Alamos National Laboratory, Los Alamos, NM 87545}

\author[JH]{D. Asthagiri}
\address[JH]{Department of Chemical and Biomolecular Engineering, John Hopkins University,  3400 North Charles Street, Baltimore, MD 21218}

\author{Surface Science, in press 2007}

\begin{abstract}
Plateaus in water adsorption isotherms on hydroxylated BeO surfaces suggest
significant differences between the hydroxylated $(100)$ and $(001)$
surface structures and reactivities.  Density functional theory
structures and energies clarify these differences.  Using relaxed
surface energies, a Wulff construction yields a  prism crystal shape
exposing long $(100)$ sides and much smaller $(001)$ faces.  This is
consistent with the BeO prisms observed when beryllium metal is
oxidized.  A water oxygen atom binds to a single surface beryllium  ion
in the  preferred adsorption geometry on either surface. The water
oxygen/beryllium bonding is stronger on the surface with greater
beryllium atom exposure, namely the less-stable $(001)$ surface.  Water/beryllium
coordination facilitates water dissociation.  On the $(001)$ surface,
the dissociation products are a hydroxide bridging two beryllium ions and a
metal coordinated hydride with some surface charge depletion.  On the
$(100)$ surface, water dissociates into a hydroxide ligating a Be atom and a proton
coordinated to a surface oxygen but the
lowest energy water state on the $(100)$ surface is the undissociated
metal-coordinated water.  The $(100)$ fully hydroxylated surface
structure has a hydrogen bonding network which facilitates rapid proton
shuffling within the network.  The corresponding $(001)$ hydroxylated
surface is fairly open and lacks internal hydrogen bonding.  This
supports previous experimental interpretations of the step in water
adsorption isotherms.  Further, when the $(100)$ surface is heated to
$1000$~K, hydroxides and protons associate and water desorbs.  The more
open $(001)$ hydroxylated surface is stable at $1000$~K.  This is
consistent with the experimental disappearance of the isotherm step when
heating to $973$~K.
\end{abstract}

\end{frontmatter}

\section{Introduction}
\label{Intro}

As a result of its low density, high melting point, nuclear properties,
and stability, beryllium is used in aerospace,
ceramics, electronics, nuclear energy and defense, metal
recycling industries, dentistry, and even sporting goods \cite{Stonehouse1986}.  Up to
temperatures of $1534$~K and pressures of 12~GPa, beryllium crystal is
hexagonal closed packed (HCP) \cite{Vijayakumer1984,Ming1984}.  
In air, beryllium oxide (BeO) forms on surfaces of beryllium
metal\cite{Adams1969,Hoover1989}.  BeO has been identified as an
important component of beryllium aerosol exposures that lead to chronic
beryllium disease (CBD), an incurable disease primarily of the pulmonary
region of the lung\cite{Stefaniak2004}.   Beryllium machining is likely
to produce beryllium aerosol particles coated with BeO. Aerosol
particles deposited in the pulmonary region of the lung are found in
phagolysosomes within macrophage cells.  In the low pH (about 4.8) of
these macrophage vacuoles beryllium particles have been observed to
undergo dissolution\cite{Sturgill1994}. Studies in rats suggest that
beryllium oxide can cause internal damage to macrophages which then
aggregate and form granulomas \cite{Sanders1975}. Animal studies
\cite{Haley1989,Haley1994,Finch1996} as well as dissolution experiments
\cite{Stefaniak2005,Day2005} suggest that BeO calcined at $773$~K (low
fired) dissolves faster and leads to greater granuloma formation than
BeO calcined at $1273$K (high fired).  More soluble beryllium compounds
such as beryllium sulfate led to increased toxicity \cite{Finch1988}.
Understanding the solubilities of different beryllium compounds is
important in determining risk factors for CBD as well as understanding
the pathway from inhalation to the disease.  Sutton and Burastero have
studied how beryllium speciation and solubility is influenced by pH,
beryllium concentration, and the composition of the dissolving
fluid\cite{Sutton2003}.  Since the mechanism of dissolution depends on
the  surface structure, characterizing common BeO surfaces and their
reactivity is a useful first step.  In the present study, we
characterize the typical surfaces observed when BeO is formed by
oxidizing beryllium metal and their hydroxylation. Hydroxylation can occur upon exposure to water vapor in air \cite{Morimoto1974,Miyazaki1985} prior to inhalation or at the latest when the BeO comes into contact with water in neutral pH lung tissue prior to ingestion by a macrophage.  In a later study, we will consider the effect of the acidic environment within the phagolysosomes of the macrophage cell on BeO dissolution.

Beryllium oxide (BeO) is a Wurtzite crystal up to pressures of
137~GPa \cite{Chang1984}.  LEED experiments suggest that a six to seven
monolayer BeO $(001)$ film grows on top of Be$(001)$.  A new $p(2\times 2)$ LEED pattern appears which suggests a surface
reconstruction \cite{Fowler1984.283}.  Crystal growth experiments on both
Be/BeO and the analogous Zn/ZnO system suggest that when the metal is burned, needles
of the oxide appear.  These needles grow into prisms.  The top face of
the prisms is the $(001)$ face while the side of the prisms is the
$(100)$ face.  Water adsorption studies on BeO crystals at temperatures
below $973$K show a plateau in the adsorption isotherms.  This plateau is
linked to fast water chemisorption on defect sites, followed by fast
water adsorption on the hydroxylated  $(001)$ and $(00\bar{1})$
surfaces, followed by a slow water adsorption on the hydroxylated
$(100)$ surface.  Hydroxylation of the $(001)$ and $(00\bar{1})$
surfaces was expected to yield fairly isolated hydroxyl groups which
could readily hydrogen bond to adsorbed water \cite{Morimoto1974,Miyazaki1985}.  In contrast,
hydroxylation of the $(100)$ surface was expected to yield a network of
hydrogen bonded hydroxyls which would attract additional
water molecules for comparatively weak absorption.

The plateau in the adsorption isotherms becomes less pronounced as the
temperature rises until it is completely gone by $973$K.  IR
spectra of the same systems show a decrease in hydroxylation with
increasing temperature until there is no OH stretch signal at
$973$K and beyond. It is suggested that surface rumpling at higher
temperatures occludes the beryllium and disrupts the hydrogen-bond
network among surface hydroxyls \cite{Morimoto1974,Miyazaki1985}.  The
temperature range below which  hydrogen-bonded
surface-hydroxyl network forms ($973$~K) is in
between the high and low fired calcined BeO temperatures
\cite{Haley1989}.   Basic chemistry suggests that the hydroxylation
structure of the different BeO surfaces plays an important role in how
the surfaces interact with water and hence the dissolution process.  The
calcined BeO and animal studies suggest that solubility is important to
the formation of granulomas and development of CBD
\cite{Stefaniak2005,Day2005,Finch1988}.

This paper compares density functional theory geometries and energies
for water adsorption, dissociation, and the resulting hydroxylation on
the $(001)$ and $(100)$ BeO surfaces.  These surfaces are exposed
when BeO grows on Be$(001)$. Understanding hydroxylation of these
surfaces is a prerequisite to understand the dissolution.  Section~\ref{Bulk} describes the 
method used to calculate bulk BeO lattice parameters.  Section~\ref{surfaces} presents the
$(001)$, $(00\bar{1})$ and $(100)$ BeO surface cuts and relaxations. 
Water adsorption and dissociation on these surfaces is discussed in
section~\ref{adsorption}.  Section~\ref{hydroxylated} describes the
hydroxylated surface structures.  Finally, in section~\ref{discussion},
our findings are discussed in relation to available experiments.

\section{Bulk}
\label{Bulk}

The Vienna Ab-initio Simulation Package
(VASP \cite{Kresse1993Thesis,Kresse1993,Kresse1996,Kresse1996.2})
implementation of density functional theory with the generalized
gradient approximation and the RPBE functional~\cite{zhang} was used for all
calculations.  The projector augmented wave (PAW) potentials were used
as supplied by Kresse and Hafner \cite{Block1994,Kresse1999}. 
Specifically, the valence states of Be:$1s^2 2s^2$ and  O: $2s^2 2p^4$
were used.  The actual occupancies are adjusted during the
self-consistent field electronic cycles while maintaining the total
number of electrons fixed.  K-point meshes were generated using the
$\Gamma-$point shift. The cutoff for the plane waves was $400$~eV.  
Electronic optimization stops when the total system energy and the band
structure energy of successive steps differs by less than $10^{-5}$~eV. 
Projection operators were evaluated in reciprocal space.  Electronic
optimization  used the the special Davidson block iteration scheme implemented in VASP\cite{davidson,liu}.  Gaussian smearing
with a width of $0.1$~eV was used.  Geometry optimizations used a
conjugate gradient method.    The geometry optimization stops when
successive energy differences are less than $10^{-4}$~eV.  Ions, cell
shape and cell volume were varied.  Precision was kept at the accurate
level to avoid wrap around errors for a very accurate bulk calculation. 
 Bulk calculations used a primitive mesh of $8 \times 8 \times 6$
k-points.  The Wurtzite primitive vectors $(\sqrt{3}a/2,-a/2,0)$,
$(0,a,0)$, and $(0,0,c)$ were used in the optimizations.  This
corresponds to a slab of dimensions $a\times a\times c$.  The $a$ and
$c$ lattice constants found were $2.71$ and $4.39$~\AA, respectively.  These values compare
well with experimental values of $2.6967(1)$ and $4.3778(1)$~\AA\cite{Reckeweg2003}. 

\section{$(001)$, $(00\bar{1})$ and $(100)$ BeO surface relaxations}
\label{surfaces}

The total cell size (slab and vacuum) and shape are kept at their bulk
values while relaxing the surfaces.  Additionally, the inner two BeO
layers are kept at the bulk geometry.  The conditions for electronic and
nuclear optimization are the same as for the bulk unless otherwise noted.

\begin{figure}[p]
\caption{\label{Surfaces001m1}(a) A $001$ cut reveals two surfaces:  The top $(001)$ surface viewed from the side features exposed Be atoms while the bottom $(00\bar{1})$ surface buries them.  The $(00\bar{1})$ surface is adjacent to the Be(001) face when the beryllium oxidizes. (b) As can be seen from a top view of the reconstructed $(001)$ surface, this lowest energy reconstructed surface has Be atom vacancies (black circles) in the flattened $(001)$ face and additional Be atoms at three fold sites (light grey spheres in the center of hexagons) on the $(00\bar{1})$ face.  In all figures, the small light grey and large dark grey spheres are Be and O, respectively.}
\begin{center}
\includegraphics[width=4cm]{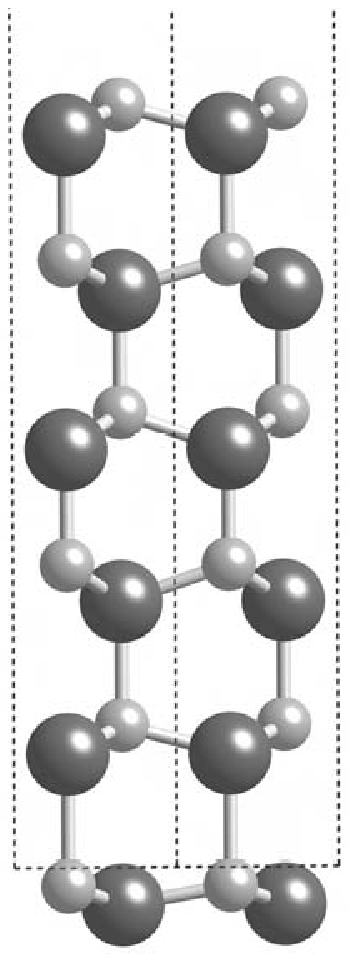}
\newline
(a)
\newline
\includegraphics[width=8cm]{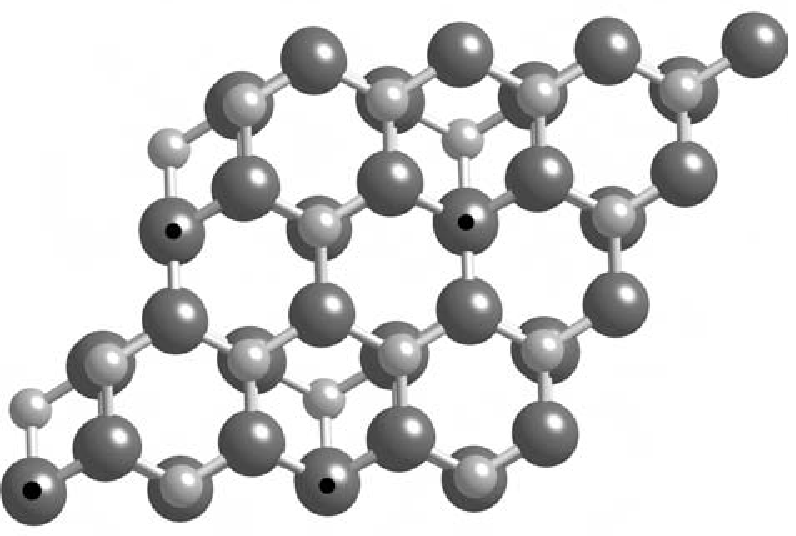}
\newline
(b)
\newline
\end{center}
\end{figure}

As shown in Figure~\ref{Surfaces001m1} (a), a cut through the $(001)$
plane results in a non-symmetric slab with a $(001)$ surface on one end
and the $(00\bar{1})$ surface at the other end.  This gives rise to a
dipole which induces a reconstruction when the surface is six to seven
layers thick \cite{Fowler1984.283}.  As can be seen from
Table~\ref{SurfaceEnergies}, a six layer slab of area $a\times a$ with
vacuum-width of
$14.76$~\AA\ is sufficient to converge surface energies. 
Throughout this paper, the width of the vacuum layer is defined as the
distance between the outermost atoms in the simulation box surfaces
along the $z$ and $x$ directions for $(001)/(00\bar{1})$ and $(100)$
surfaces, respectively.  An adatom is never considered as an outermost
surface atom.  The $(001)/(00\bar{1})$ calculations use a 
k-point mesh of $8\times 8 \times 1$ and dipole corrections.

Fowler, \emph{et~al.},  \cite{Fowler1984.283} postulated a reconstruction
of a $(001)/(00\bar{1})$ slab where $1/4$ of either cation vacancies or
adsorbed anions form on the $(001)$ surface and the same number of
adsorbed cations or anion vacancies form on the $(00\bar{1})$ side.  The
LEED pattern is consistent with the anion and cation additions appearing
in their normal lattice sites.  The six layer $a\times a$ slabs used for
the surface relaxations only contain single beryllium and oxygen ions on
each surface.  Therefore, moving $1/4$ of the cations or anions from one
surface to another requires a surface that is at least $2a \times 2a$. 
Comparing the relaxed structures of this minimum size surface and a $4a
\times 4a$ surface revealed that the minimum size was not large enough. 
Using a $2\times 2\times 1$ k-point grid on a six layer $3a \times 3a$
slab with a $19.15~$\AA~wide vacuum layer yielded a relaxed surface
energy of $0.203$ eV/\AA$^2$ which is in agreement with the relaxation
energy of the smaller surfaces with a larger grid.  This suggested that
a $2\times 2\times 1$ k-point grid is sufficient for these larger
surfaces.  Optimizations were started from: (i) a relaxed surface slab
with $1/4$ of the cations moved from the $(001)$ side to the $(00\bar{1})$ side
and (ii) a relaxed surface slab with $1/4$ of the anions from the $(00\bar{1})$
side moved to the $(001)$ side.  The lowest energy minimum in both cases
shifts the displaced ions from their normal lattice sites to binding in
three fold sites.  As seen in Table~\ref{SurfaceEnergies}, the
$p(2\times2)_i$ and $p(2\times2)_{ii}$ reconstructed surfaces have lower
surface energies than the unreconstructed $(001)/(00\bar{1})$ surfaces. 
The lowest energy reconstructed surface found is shown in
Figure~\ref{Surfaces001m1} (b).  The exposed $(001)$ surface of this structure is consistent with the LEED pattern.  However, there is discrepancy on the unexposed $(00\bar{1})$ surface.  In the experiment, ions are kept in lattice positions by the beryllium metal below.  Since our system does not have a BeO/Be interface the beryllium atoms are displaced from their normal lattice positions to three-fold sites on the $(00\bar{1})$ surface.

\begin{figure}[p]
\caption{\label{Surface100}The $(100)$ cut reveals two symmetric $(100)$ surfaces.  In this relaxed surface, the surface beryllium ions are slightly lower than the oxygens.}
\begin{center}
\includegraphics[width=8cm]{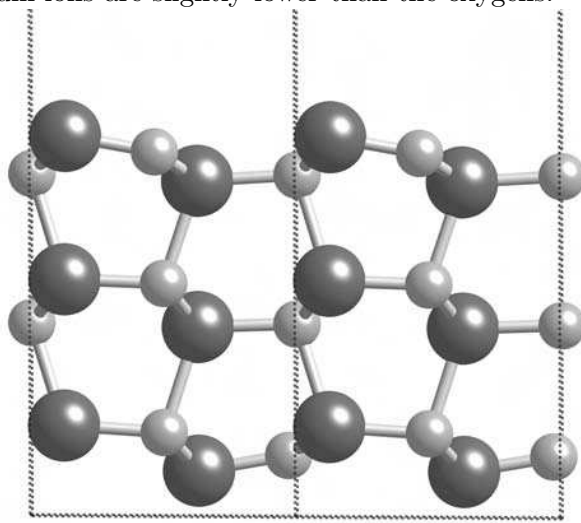}
\end{center}
\end{figure}

Table~\ref{SurfaceEnergies} also shows the surface energies of different
$(100)$ slabs using k-point grid of $1 \times 8 \times 6$.  A three
layer $a \times c$ slab with $9.87~$\AA~of vacuum is sufficient for
converging $(100)$ surface energies.  The dipole correction was
negligible for this surface.  Figure~\ref{Surface100} displays our
relaxed (100) surface.  The distortions from bulk show the same trends
as the previous calculations of Jaffe and Zapol\cite{Jaffe1997}.  For example, the distance between surface beryllium and oxygen atoms in the relaxed surface structure is $1.50 ~$\AA~in our calculations and $1.48~$\AA~in their calculation.  The same distance for atoms just below the surface is $1.67 ~$\AA~in our calculation and $1.63~$\AA~in their calculation.  In both calculations, oxygens occlude beryllium.

\section{Water adsorption and dissociation}
\label{adsorption}
A water molecule was placed on each surface in a geometry intended to
favor bonding.  After optimization of the initial structures,
neighboring geometries were sampled using molecular dynamics (MD) on the Born-Oppenheimer surface with a time step of 1~fs and temperature re-scaling of velocities every $10$ steps
at $800$~K. No equilibration was done.  This is a sampling scheme similar to forced bias\cite{berne}, smart\cite{doll}, and hybrid\cite{duane} Monte Carlo schemes  rather than a true dynamics.  The $\Gamma$ point and a $300$~eV plane wave cut off was used in this low precision sampling process.  Water dissociation occurred within 500 MD steps on the $(001)$ surface.  On the $(100)$, after many trials a dissociation was observed within 2000 MD steps.  The geometries sampled
were visualized and a low-energy subset was chosen for subsequent conjugate gradient
optimization.  The final optimizations used the earlier $600$~eV plane
wave cutoff and default precision.  $(001)/(00\bar{1})$ and $(100)$
adsorbate optimizations used $2\times2\times1$ and $1\times2\times2$
k-point meshes, respectively.

These optimizations revealed adsorbed and dissociated water
configurations.  Water adsorbs on the $(100)$ surface both with a water
oxygen bonded directly to a surface beryllium ion \emph{and} with a
hydrogen bond to a surface oxygen atom. After several trials starting from a bound water, water molecule dissociation was seen in a high-temperature trajectory.  Other times, the water molecule desorbed.  On the $(001)$ surface which exposes beryllium, only surface Be-OH$_2$ bonds were observed.  Water dissociated readily on the unreconstructed $(001)$ surface.

When a trajectory produced water molecule dissociation, the Nudged Elastic Band (NEB)\cite{Henkelman2000.2} method was used to find the transition pathway. 
Six linearly interpolated images between adsorbed and dissociated water
were used.  When NEB suggested intermediate minima, the minima were
refined and a new shorter path NEB calculation was done.  Once NEB was
converged, climbing NEB \cite{Henkelman2000.1} was implemented to refine
the transition state. Table~\ref{WaterEnergies} shows the system size
and binding energies of water in the adsorbed, transition state, and
dissociated forms on $(001)$ and $(100)$ surfaces.  All these energies
are relative to the corresponding bare relaxed surface plus an isolated
water molecule.  The dissociated product on the $(001)$ surface is $2.57$~eV lower in energy than the undissociated bound water.  The dissociation barrier is only $0.12$~eV.  In contrast, the dissociation barrier on the $(100)$ surface is  $0.57$~eV and the dissociated product is $0.36$~eV higher in energy than the undissociated bound water.  This explains why water dissociates more readily on the $(001)$ surface than on the $(100)$ surface.  The size of the $(001)$ surface needed for reconstruction made water dissociation studies on it computationally prohibitive.

\begin{figure}[p]
\caption{\label{Water001} Water can dissociate on the $(001)$ surface by first binding to a beryllium on the surface (a), stretching an OH bond in the direction of a beryllium, and finally reaching a dissociated state (b).  The transition state looks similar to (a) and hence is not shown.  This dissociated state involves a hydroxyl, a hydride, and some electron density reductions on the surface.  The white spheres are hydrogens.  Multiple paths to this type of dissociation are possible.  This is a representative one.}
\begin{center}
\includegraphics[width=6cm]{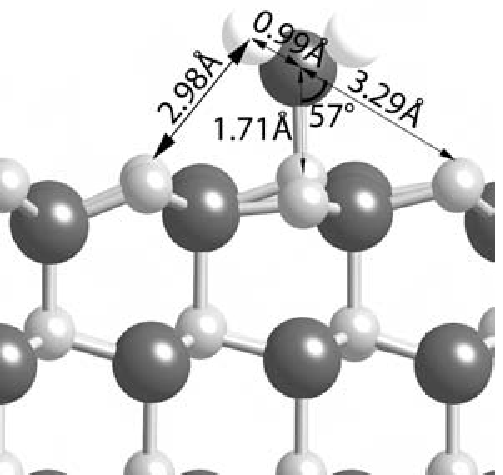}
\newline
(a)
\newline
\includegraphics[width=6cm]{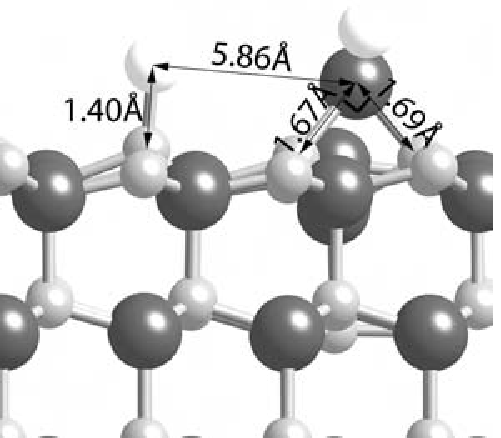}
\newline
(b)
\newline
\end{center}
\end{figure}

\begin{figure}[p]
\caption{\label{Water100} Water can dissociate on the $(100)$ surface by first binding to a beryllium on the surface (a), going through a transition state in which the water hydrogen bonds to an adjacent oxygen (b), and finally reaching a dissociated state (c).  This dissociated state involves a proton and a hydroxide.}
\begin{center}
\includegraphics[width=6cm]{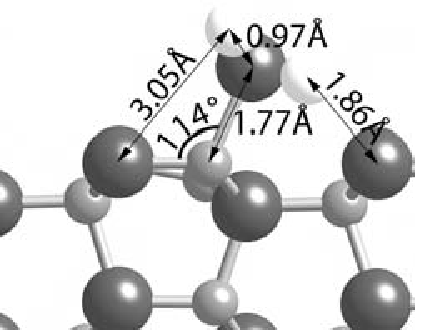}
\newline
(a)
\newline
\includegraphics[width=6cm]{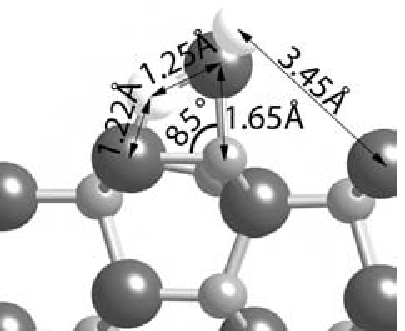}
\newline
(b)
\newline
\includegraphics[width=6cm]{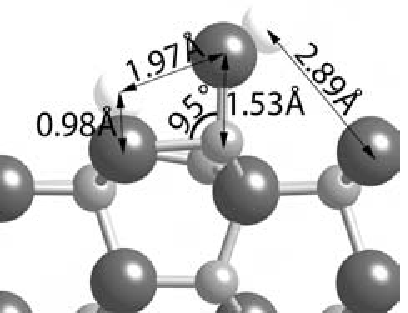}
\newline
(c)
\newline
\end{center}
\end{figure}

Figure~\ref{Water001} shows the corresponding water adsorption and dissociated geometries on the $(001)$ surface.  The transition state looks similar to the bound water minimum and hence is not shown. 
Figure~\ref{Water100} shows the water
adsorption, transition state, and dissociated geometries found for the
$(100)$ surface.  Water dissociation on the
$(001)$ surface differed significantly from dissociation on $(100)$.  On
the $(001)$ surface Be-OH$_2$ loses a hydrogen to a nearby
beryllium atom.  The hydroxide shifts to coordinate with a second
beryllium.   Bader analysis \cite{Henkelman2005} of the electron charge
density suggests that a hydride and a hydroxide have been formed, as 
the hydrogen and hydroxide charges are $-0.71$ and $-1.06$,
respectively.  To compensate, there is a lower electron density among
some of the other surface atoms.  When more waters are added, further
hydroxylation is possible.  \emph{In this case, dynamic runs show hydrogen atoms
combining,  leaving as hydrogen gas, thus giving a direct demonstration of
partial oxidation of the beryllium.}  In contrast, on the $(100)$
surface, water dissociates into a hydroxide coordinated to a single
surface beryllium and a proton which binds to a surface oxygen.  Bader
charge density analysis of the dissociated water on the $(100)$ surface
reveals a $-1.01$ charge on the hydroxide and a $+1.00$ charge on the
proton.  As can be seen in Figure~\ref{Water100}, the dissociated
configuration permits hydrogen bonding between the hydroxide oxygen and
the adjacent proton.

\section{Hydroxylated surface structure} \label{hydroxylated}

\begin{figure}[p]
\caption{\label{Hydroxylation} (a) Hydroxylated $(001)$ surface viewed from the $<001>$ direction. (b) Hydroxylated $(100)$ viewed from the $<100>$ direction.}
\begin{center}
\includegraphics[width=8cm]{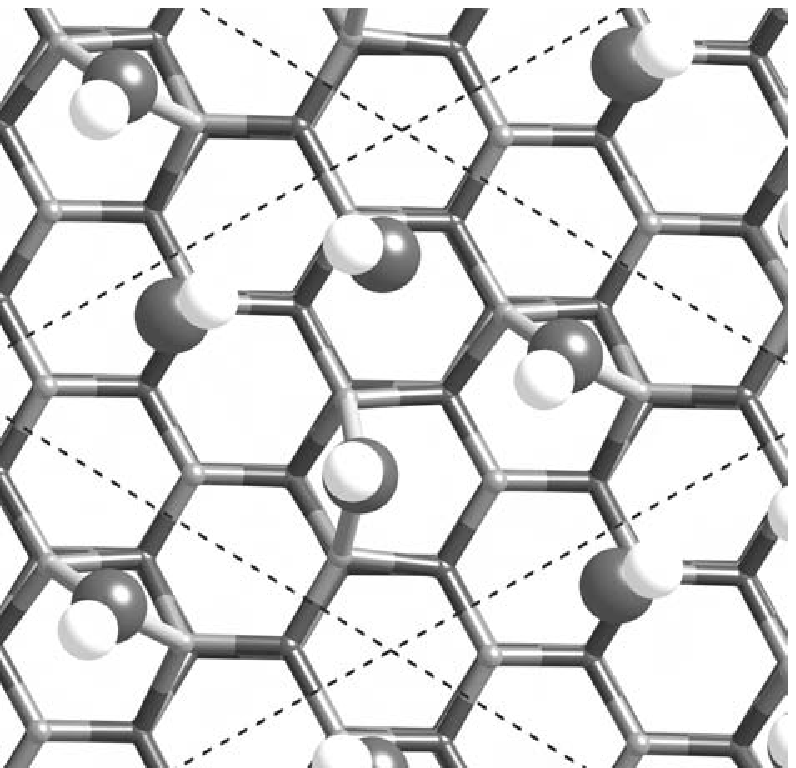}
\newline
(a)
\newline
\includegraphics[width=8cm]{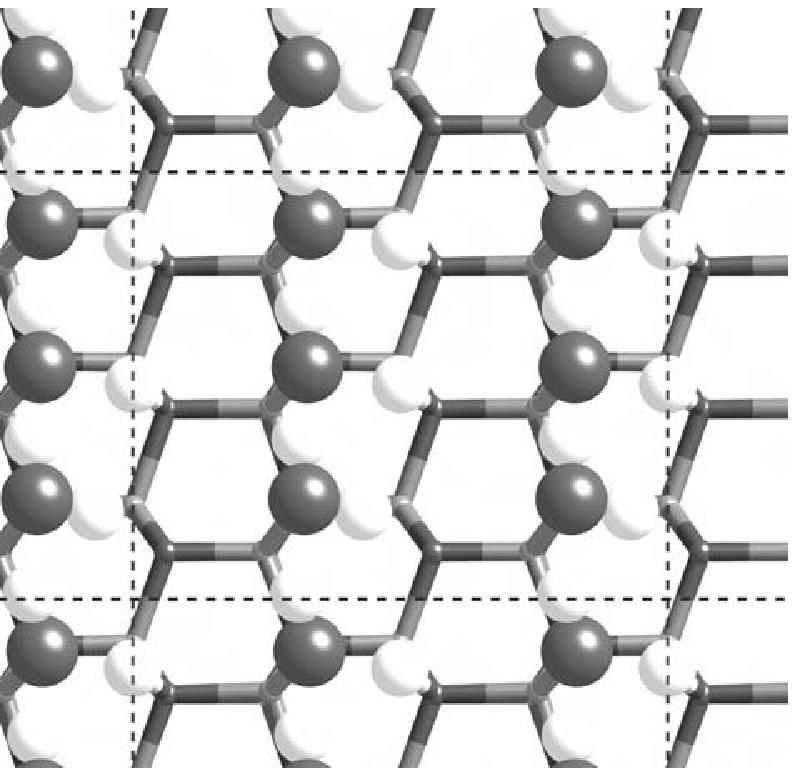}
\newline
(b)
\newline
\end{center}
\end{figure}

Surfaces were fully hydroxylated as suggested by the dissociation
patterns seen in the molecular dynamics sampling.  The hydroxylated
surfaces were optimized using the $\Gamma$ point.  From this
configuration, a sampling molecular dynamics was started. Configurations
from this run were then optimized.  There are in principle many
orientations possible for all the hydroxide combinations and this is by
no means an exhaustive minima search.  Nevertheless,
Figures~\ref{Hydroxylation} (a) and (b) show a representative
hydroxylation structure for the $(001)$ and $(100)$ surfaces,
respectively.  The $(001)$ surface tended to form doubly-coordinated
metal hydroxide structures, and to lose hydrogen gas leading to a
sparsely hydroxylated $(001)$ surface.  Additionally, once some
doubly-coordinated beryllium ions are produced, the chance for further
hydroxylation decreases.  Hence, the surface hydroxylation was not
homogeneous as seen in the sample configuration shown in
Figure~\ref{Hydroxylation} (a).  All the nonbonded oxygen/hydrogen
distances there are larger than $3~$\AA, \emph{i.e.}, longer than traditional
hydrogen bonds.

In contrast, the tendency of the $(100)$ surface to dissociate water
molecules into a hydroxide and proton on adjacent surface beryllium and oxygen ions leads to a denser hydroxylation.  On this hydroxylated
surface, adjacent non-bonded oxygens and hydrogens have distances in the
range of $1.6$ to $1.9$~\AA.  In addition, sometimes a row of adjacent
hydroxides and protons associated to beryllium bonded water, and
occasionally more than one row associated in that way.  This is
consistent with the hydrogen bonded network postulated in
references~\cite{Morimoto1974,Miyazaki1985} to explain the step in water
adsorption data.

Heating the $(100)$ hydroxylated surface to $1000$~K lead to proton
shuffling and association of hydroxide ligands with protons followed by water
desorption of several water molecules.  The same heating of the $(001)$
hydroxylated surface left the hydroxylated structure unchanged.

\section{Concluding discussion}
\label{discussion}

Bulk lattice parameters agree with experiment\cite{Reckeweg2003}.  Beryllium oxide prisms
with prominent $(100)$ sides and smaller $(001)$ tops grow on
Be$(001)$ \cite{Morimoto1974,Miyazaki1985} in the laboratory.  Consistent
with this, a Wulff construction \cite{Abramowski1999,Wulff1901} based on
the surface energies in Table~\ref{SurfaceEnergies} suggests that BeO
crystals exposing $(100)$, $(001)$ and $(00\bar{1})$ faces exhibit a
prism shape with large $(100)$ faces and small $(001)$ and $(00\bar{1})$
faces.  Some synthetic crystals which expose these faces have the same
prism shape \cite{Austerman1963}.  When BeO is grown on Be$(001)$, the
BeO$(00\bar{1})$ face is hidden by the Be$(001)$.  There are also
alternative ways to make BeO crystals which expose additional phases and
exhibit different crystal patterns \cite{Austerman1964,Austerman1977}.

In agreement with LEED experiments \cite{Fowler1984.283} where thin
layers of BeO are grown on Be(001), a $p$(2x2) lower energy
reconstruction of the BeO$(001)$ is found.  Nevertheless, our lowest energy
reconstruction has beryllium in three fold sites rather than at the
standard lattice positions.  This is likely a result of not having a
Be$(001)$ substrate attached to the BeO$(00\bar{1})$ surface in our simulations.  The $(001)$ surface reconstruction is consistent with the experiment.

Water adsorption and dissociation reveals a reactive
unreconstructed $(001)$ surface consistent with preferred BeO growth in
the direction of the $(001)$ surface normal making more $(100)$ surface.
 The preferred state of water on the $(001)$ surface is in the
dissociated form.  In contrast, on the $(100)$ surface, water
prefers to physi-sorb.  The size of the $(001)$ reconstructed simulation cells prohibited detailed study of water/surface reactivity on these systems.  Since the reconstructed  flatter $(001)$ surface has beryllium only slightly higher than oxygens, water surface reactivity is expected to decrease relative to the unreconstructed surface.   However, water reactivity should still be greater than for the $(100)$ surface where surface berylliums are below surface oxygens.

Extensive hydroxylation of the unreconstructed $(001)$ surface does not
allow for much internal hydrogen bond formation.  In contrast, hydroxylation
of the $(100)$ surface lends itself to a hydrogen bonded network.  This might lead to a slower water adsorption until a new water layer capable of hydrogen bonding to additional water is added.  This 
may explain the step in the adsorption isotherms. 
Finally, the hydrogen bonded network in the hydroxylated $(100)$ surface allows
for significant proton shuffling aiding association of hydroxide and
proton into water.  Desorption follows heating to $1000$~K.  This is
consistent with the disappearance of the adsorption step and IR OH
stretch signal following exposure to this
temperature \cite{Morimoto1974,Miyazaki1985}.  Nevertheless, in our
simulations, the surface does not appear to have a permanent change.

The presence of reactive surfaces such as the (001) here
might explain the high initial rates observed in
dissolution studies\cite{Finch1989}, although dissolution of BeO in
aqueous media is acid catalyzed \cite{Furrer1986}.  Therefore in sufficiently
acidic solutions equilibration of surface structures should probably be
expected after an initial rate. Then a hydroxylated surface
may be an obligatory intermediate for the molecular mechanism of
dissolution.  Surface dehydroxylation at high temperatures might be
indicative of annealing that could explain the slower dissolution of
high fired beryllium oxide compared to low fired beryllium oxide.

\section{Acknowledgments}

We would like to thank Aleksandr B. Stefaniak and Ronald C. Scripsick
for useful discussions.  This work was carried out under the
auspices of the National Nuclear Security Administration of the U.S.
Department of Energy at Los Alamos National Laboratory under Contract
No. DE-AC52-06NA25396.  



\bibliography{BeOLibrary}
\bibliographystyle{unsrt}






\newpage
\listoftables
\listoffigures
\newpage

\clearpage
\begin{table}
\caption{\label{SurfaceEnergies} Surface energies per unit surface area.  Vacuum length is defined by the distance from the outermost atom on one surface to the outermost atom on the other surface along the z or x axes for the $(001)/(00\bar{1})$, denoted $(001)$, and $(100)$ surfaces, respectively.  An adatom is never considered as an outermost surface atom.  $p(2\times 2)_i$ and $p(2\times 2)_{ii}$  are the reconstructions moving:  (i) $1/4$ cations moved from the $(001)$ side to the $(00\bar{1})$ side
and (ii)  $1/4$ anions from the $(00\bar{1})$ side moved to the $(001)$ side.  See Section~\ref{surfaces} for more details.   See Figures~\ref{Surfaces001m1} (a) and~\ref{Surface100} for pictures of the relaxed but unreconstructed $(001)$ and $(100)$ surfaces.   Figure~\ref{Surfaces001m1}(b) shows the $p(2\times 2)_i$ reconstructed structure.} 
\begin{center}
\begin{tabular}{|lrr|r|r|r|}
\hline
 Surface&\multicolumn{2}{l|}{Slab Dimensions}&Vacuum&\multicolumn{2}{l|}{Surface Energy (eV/\AA$^2$)}\\
\cline{5-6}
 &Area&Layers&Height (\AA) &Unrelaxed&Relaxed\\
\hline\hline
Unreconstructed $(001)$:&$a\times a$&6&14.76&0.220&0.194\\
&$a\times a$&6&19.15&0.218&0.193\\
&$a\times a$&8&19.10&0.221&0.198\\
$p(2\times 2)_i$  & $4a\times 4a$&6&19.39&&0.090\\
$p(2\times 2)_{ii}$ & $4a\times 4a$&6&18.96&&0.113\\
$(100):$&$a \times c$&3&9.87&0.120&0.069\\
&$a \times c$&4&12.58&0.120&0.070\\
\hline
\end{tabular}
\end{center}
\end{table}

\clearpage
\begin{table}
\caption{\label{WaterEnergies} The system size and binding energies of water in the adsorbed, transition state, and dissociated forms on $(001)$ and $(100)$ surfaces.   All these energies are relative to the corresponding bare relaxed surface plus an isolated water molecule.  The corresponding structures are shown in Figures~\ref{Water001} and~\ref{Water100}, respectively.  Moving the hydride in the $(001)$ structure lowers the energy further.}
\begin{center}
\begin{tabular}{|lrr|r|r|r|r|}
\hline
 Surface&\multicolumn{2}{l|}{Slab Dimensions}&Vacuum&\multicolumn{3}{l|} {Binding Energy (eV)}\\
\cline{5-7}
 &Area&Layers&Height (\AA) &Adsorption&Transition State&Dissociation\\
\hline\hline
$(001)$:&$3a\times 3a$&6&19.15&-1.28&-1.16&-3.85\\
$(100)$:&$3a\times 2c$&3&26.12&-0.75&-0.18&-0.39\\
\hline
\end{tabular}
\end{center}
\end{table}

\end{document}